\documentclass[fleqn,usenatbib]{mnras}

\usepackage[T1]{fontenc}

\DeclareRobustCommand{\VAN}[3]{#2}
\let\VANthebibliography\thebibliography
\def\thebibliography{\DeclareRobustCommand{\VAN}[3]{##3}\VANthebibliography}

\usepackage{graphicx}
\usepackage{subfigure}	
\usepackage{amsmath}	
\usepackage{amssymb}	
\usepackage{color}
\usepackage{changes}
\usepackage{cjhebrew}

\title[Change of solar wind velocity revealed by FAST]{Linear change and minutes variability of solar wind velocity revealed by FAST}

\author[Li-Jia Liu et al.]{
Li-Jia Liu,$^{1,2,3}$
Bo Peng,$^{1,6}$\thanks{E-mail: \href{mailto:pb@nao.cas.cn}{pb@nao.cas.cn}}
Lei Yu,$^{1,3}$\thanks{E-mail: \href{mailto:yulei@nao.cas.cn}{yulei@nao.cas.cn}}
Bin Liu,$^{1,6}$\thanks{E-mail: \href{mailto:bliu@nao.cas.cn}{bliu@nao.cas.cn}}
Ji-Guang Lu,$^{1}$
Ye-Zhao Yu,$^{4}$
Hong-Wei Xi,$^{1,3}$
\and Ming Xiong,$^{3,5}$
O. Chang,$^{7}$
\\
\\
$^{1}$CAS Key Laboratory of FAST, National Astronomical Observatories, Chinese Academy of Sciences, Beijing 100101, China\\
$^{2}$State Key Laboratory of Space Weather, Chinese Academy of Sciences, Beijing 100190, China\\
$^{3}$University of Chinese Academy of Sciences, Beijing, 100049, China\\
$^{4}$School of Physics and Electronics, Qiannan Normal University for Nationalities, Duyun, 558000, China\\
$^{5}$National Space Science Center, Chinese Academy of Sciences, Beijing 100190, China\\
$^{6}$Hebei Key Laboratory of Radio Astronomy Technology, Hebei 050081, China\\
$^{7}$RAL Space, United Kingdom Research and Innovation - Science and Technology Facilities Council - Rutherford Appleton Laboratory, \\
Harwell Campus, Oxfordshire, OX11 0QX, UK
}

\date{Accepted XXX. Received YYY; in original form ZZZ}

\pubyear{2015}

\begin{document}
\label{firstpage}
\pagerange{\pageref{firstpage}--\pageref{lastpage}}
\maketitle

\begin{abstract}

Observation of Interplanetary Scintillation (IPS) provides an important and effective way to study the solar wind and the space weather. A series of IPS observations were conducted by the Five-hundred-meter Aperture Spherical radio Telescope (FAST). The extraordinary sensitivity and the wide frequency coverage make FAST an ideal platform for IPS studies. In this paper we present some first scientific results from FAST observations of IPS with the $L$-band receiver. Based on the solar wind velocity fitting values of FAST observations on September 26-28, 2020, we found that the velocity decreases with increasing frequency linearly, which has not yet been reported in literature. And we have also detected a variation of solar wind velocity on a timescale of 3-5 minutes, which imply the slow change of the background solar wind, a co-existence of high- and low-speed streams, or a reflect of the quasi-periodic electron-density fluctuations.

\end{abstract}



\section{Introduction}

Interplanetary scintillation (IPS) is caused by the density irregularities in the solar wind plasma\citep{hewish1964interplanetary}. Observations of IPS of compact radio sources can be used to study the solar wind and the angular structures of the scintillating radio sources \citep{hewish1969radio, 1972JGR....77.4602A, manoharan1990determination}.
Compared with in-situ measurement provided by spacecraft, ground based observations of IPS can give information over all the heliographic latitudes and cover a wide range of solar elongations from the Sun \citep{1967Natur.213..343D}. Therefore, it provides an efficient way to study the heliospheric physics.\\

After IPS phenomenon was discovered in the 1960s \citep{1964Clarke, hewish1964interplanetary}, many instruments have been dedicated to the observations of IPS, such as the Ooty Radio Telescope \citep{manoharan1990determination}, the multi-station system of the Institute for Space-Earth Environmental Research \citep{asai1995multi}, the Mexican Array Radio Telescope \citep{2010SoPh..265..309M} and the Puschino radio telescope \citep{2003A&A...403..555A}. After decades of studies, the results deduced from the observations of IPS can be used to form the three-dimensional (3D) distribution sky map for solar wind velocity and electron density \citep{1998JGR...10312049J, 2019AGUFMSH43D3362J, 1998JGR...103.1981K}, and can also be used to forecast the arrival time of coronal mass ejection to the Earth environment \citep{2003AGUFMSH42C0548T, 2006SoPh..235..345M, 2019EP&S...71...39I, 2021EP&S...73....9I}. \\

The Five-hundred-meter Aperture Spherical radio Telescope (FAST), which is located in Guizhou Province of China, was completely constructed in 2016. The high sensitivity and wide frequency range of FAST make it powerful to carry out observations of IPS. \\

This paper is laid out as follows: IPS observations with FAST are presented in Section 2. The data reduction and observation results are described in Section 3. Sections 4 and 5 present the discussions and concluding remarks, respectively.\\

\section{Observation of IPS with FAST}

FAST has an aperture diameter of 500\,m and an illuminated aperture of 300\,m. As the largest single-dish radio telescope in the world, FAST has not only a high sensitivity, but also a high time and a high frequency resolution, which makes it the most powerful telescope to detect weak or short timescale signals \citep{2000pras.conf...25P}. \\

During the commissioning, the Ultra-Wideband (UWB) receiver was installed in the feed cabin between September, 2016 and July, 2018. This receiver covers a wide frequency range of 1.35\,GHz from 270 to 1620\,MHz. The huge frequency coverage allows it to carry out the single station single frequency (SSSF) and single station dual-frequency (SSDF) analysis modes observations of IPS simultaneously, which makes it possible to compare the solar wind velocities deduced from the two modes. Based on which, an optimized model-fitting method for SSSF analysis mode was developed and the frequency difference of the SSDF analysis mode were discussed. A pilot observation of IPS has been carried out, which was presented in an earlier paper \citep{2021MNRAS.504.5437L}.\\

Since July 2018, the 19-beam receiver at $L$-band was equipped on FAST. The frequency coverage of the receiver is from 1.05 to 1.45\,GHz with a system temperature of 20\,K \citep{2020Innov...100053Q}.\\

IPS phenomenon is strong in the area close to the Sun, but is weak in most of the interplanetary space. Therefore, the IPS study requires high sensitive observations made by large telescopes like FAST. The lower frequencies are useful for studying the regions that are further away from the Sun, and the higher frequencies are useful for those very close to the Sun \citep{1969ApJ...155..645C}. The frequency coverage of the $L$-band receiver on FAST is capable to probe as close as 18 solar radius from the Sun \citep{2007ChJAA...7..712Z}. \\

The FAST telescope has a digital acquisition back-end, which is the reconfigurable open-architecture computing hardware-version 2 (ROACH2) board, developed by the Collaboration for Astronomy Signal Processing and Electronics Research (CASPER). When performing observations of IPS, the data recorded by the central beam of the 19-beam $L$-band receiver was splitted into 4096 channels. The wide frequency coverage and the high frequency resolution of the FAST $L$-band receiver allow it to perform a dynamic spectrum of the IPS observations, which can be used to study the variations and microstructures of the solar wind \citep{1984Natur.312..251C, 2013SoPh..285..127F}. \\

Radio source 3C\,286 was chosen as a target source, which has also been observed during September 26-28, 2020. The solar elongation of 3C\,286 in three days was $37^{\circ}$, which was in the weak scattering regime. \\

The central-beam of the 19-beam receiver was used for the observations of IPS. The original sampling interval recorded by the digital back-end was 0.2\,ms. When carrying out the IPS data reduction, a new time interval with integration time of 20\,ms was generated by performing summation over every 100 raw sampling points along time. The telescope tracking on our target 3C\,286 stayed for 30, 60 and 30 minutes respectively, as summarized in Table~\ref{tab:1},where observing dates, on-source and off-source time are listed in columns 1 to 3.\\

\begin{table}
	\centering
	\caption{Summary of observations with FAST.}
	\label{tab:1}
	\begin{tabular}{lccccr} 
		\hline
        $Dates$ & on-source time & off-source time \\
          & (min) & (min)  &  \\
		\hline
		Sept. 26, 2020 & 30& 5\\
        Sept. 27, 2020 & 60& 5\\
        Sept. 28, 2020 & 30 & 5\\
        \hline
	\end{tabular}
\end{table}

\section{Data Reduction and Results}

In this Section, the data reduction method and power spectrum results are presented, including the mitigation of radio frequency interferences (RFI) mitigation, forming and model-fitting the SSSF mode power spectrum.

\subsection{Data Reduction}
\label{sec:Data Reduction} 

The raw data recorded by the $L$-band receiver is partially contaminated by RFIs in some channels, which should be removed. The whole frequency band is divided into multi-channels, which allows us to locate the RFI precisely. The steps are briefly described below:\\ \\
1. The original data recorded by FAST back-end was integrated to 20\,ms each point by performing summation every 100 raw sampling data along time;\\
2. RFI channels are identified and removed by the method of random sample consensus (RANSAC), which has been described in \citet{2021MNRAS.504.5437L};\\
3. The mean power spectrum is calculated for a certain time interval, usually of 50\,s;\\
4. The mean power spectrum is subtracted to remove the systematic instrumental response between different frequencies.\\

\begin{figure}
    \subfigure[]{
    \includegraphics[width=0.48\columnwidth]{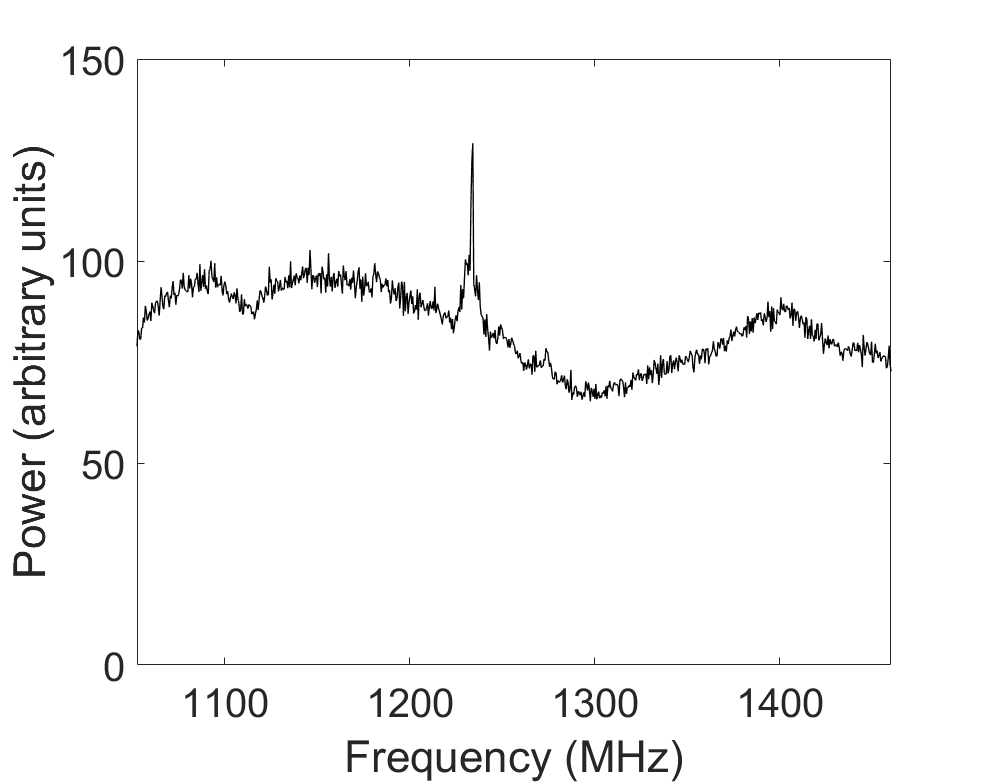}
    }
    \subfigure[]{
    \includegraphics[width=0.48\columnwidth]{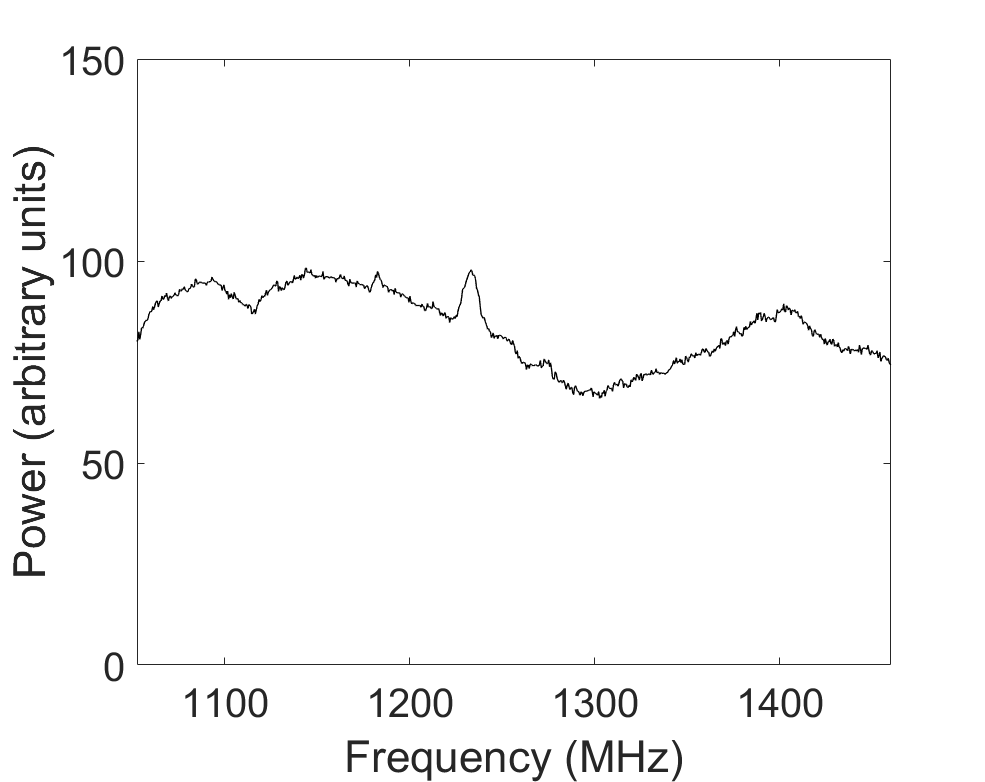}
    }

    \subfigure[]{
    \includegraphics[width=0.48\columnwidth]{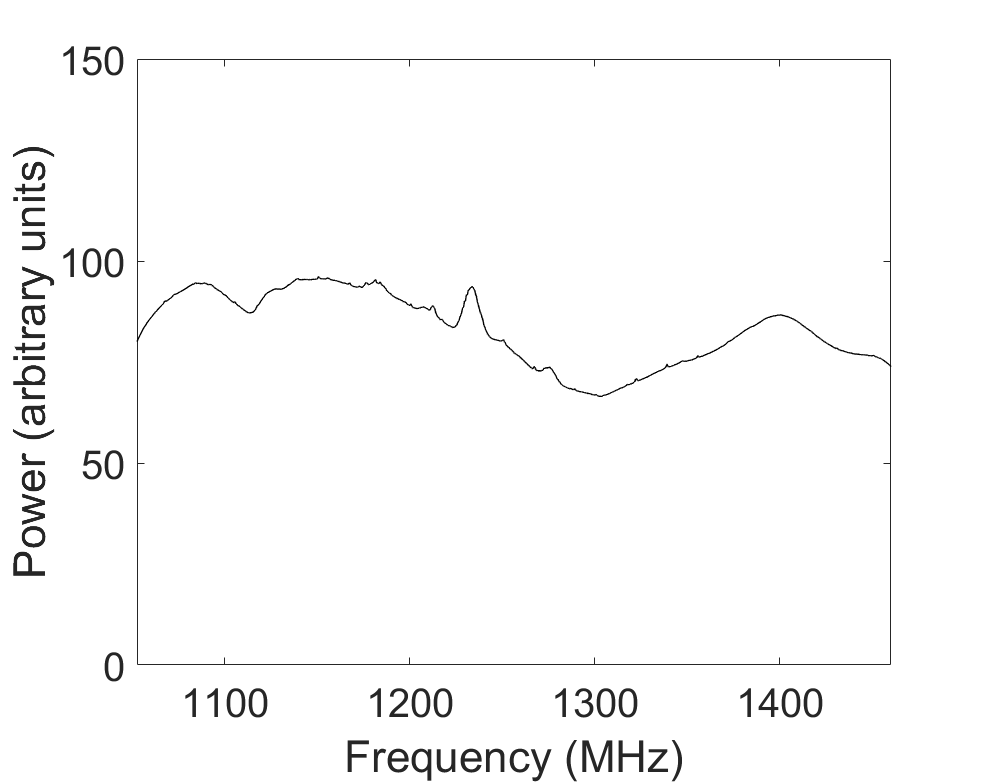}
    }
    \subfigure[]{
    \includegraphics[width=0.48\columnwidth]{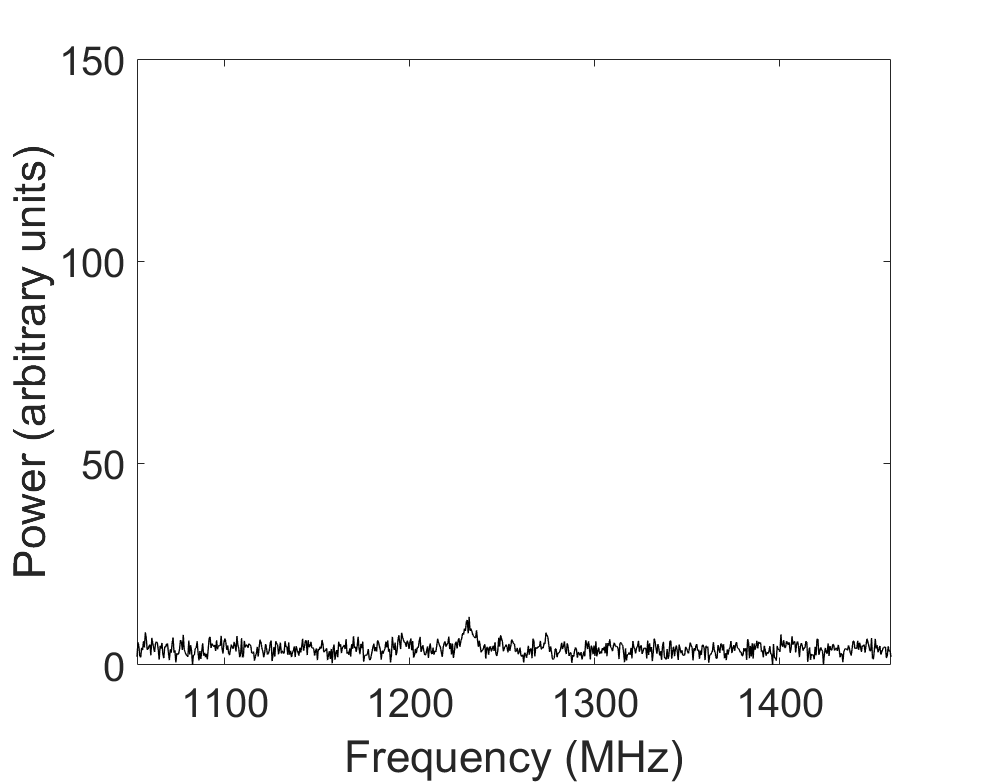}
    }
    \caption{The process of data reduction. The target source was 3C\,286 which was observed on September 27, 2020. (a)-(d) shows the power spectrum of raw data, the spectrum after removing the RFI by RANSAC, the mean spectrum and the spectrum after subtracting the mean of different frequencies respectively.}
    \label{data_process}
\end{figure}

Fig.~\ref{data_process} presents an example of the RFI removing process. 3C\,286 was chosen as a target source for FAST observation on September 27, 2020. Fig.~\ref{data_process} (a), (b), (c) and (d) show the spectrum of integrated raw data, the spectrum after removing the RFI by RANSAC, the 50\,s mean spectrum and the spectrum which has been removed by subtracting Fig.~\ref{data_process} (c). It can be seen that after the RFI mitigation, the interferences throughout the frequency bandpass had been obviously removed.\\

The data set after RFI removal was then integrated on frequency. The original sub-band frequency span was 0.12\,MHz, when carrying out IPS data processing, the sub-band will be integrated to 10\,MHz to form a new data set. Then the new time series with a bandwidth of 10\,MHz and an integration time of 20\,ms will be used for further data reduction. The SSSF mode data processing is illustrated below: \\ \\
1. The new time series is divided into blocks of length of 512 data points corresponding to a duration time of about 10\,s. Then each block is subtracted by its own average; \\
2. The data points of each block from step 1 is smoothed by using Hanning window. The fast Fourier transform is then applied to get power spectra; \\
3. The spectra of 8 sub-sets are averaged and normalized to form the SSSF mode power spectrum. \\

\subsection{Power Spectrum Results}

In our IPS studies, the solar wind velocity was deduced from a model-fitting method. The fitting parameters include 1) solar wind velocity ($V$); 2) the anisotropic axial ratio (AR) {\citep{1998JGR...103.6571Y}}; 3) the spectral index of electron density fluctuation ($\alpha$) and 4) the size of source ($\theta$). \\

For SSSF analysis mode, in the weak scattering regime, equation~(\ref{eq:1}) gives the temporal power spectrum $P(f)$ with $C=(2\pi r_e\lambda)^2$ which depends on the observing wavelength $\lambda$, $Z$ is the distance to the scattering screen, $V_p$ is the solar wind velocity projected on to the plane perpendicular to the direction of line-of-sight $z$, and $r_e$  is the electron radius. The spectrum of the electron-density fluctuations $\Phi_{ne}\propto q^{-\alpha}$, where $q=\sqrt{(q_x^2+q_y^2)} $  is the wave number. $f=\frac{q_{x}V_{p}}{2\pi}$ is the temporal frequency, $F_{diff}=4sin^2(\frac{q^2z\lambda}{4\pi})$ is the Fresnel propagation filter and $F_{source}=\exp (-(\frac{qz\theta}{2.35})^2)$ is the squared modulus of the source visibility \citep{manoharan1990determination, 2015SoPh..290.2539M}.\\

\begin{equation}
\centering
    P(f)=C\int ^{Z} _{-Z}\frac{1}{V_p}dz\int^{+\infty}_{-\infty}\Phi_{ne}F_{diff}F_{source}dq_y.
	\label{eq:1}
\end{equation}

\begin{figure}
    \includegraphics[width=\columnwidth]{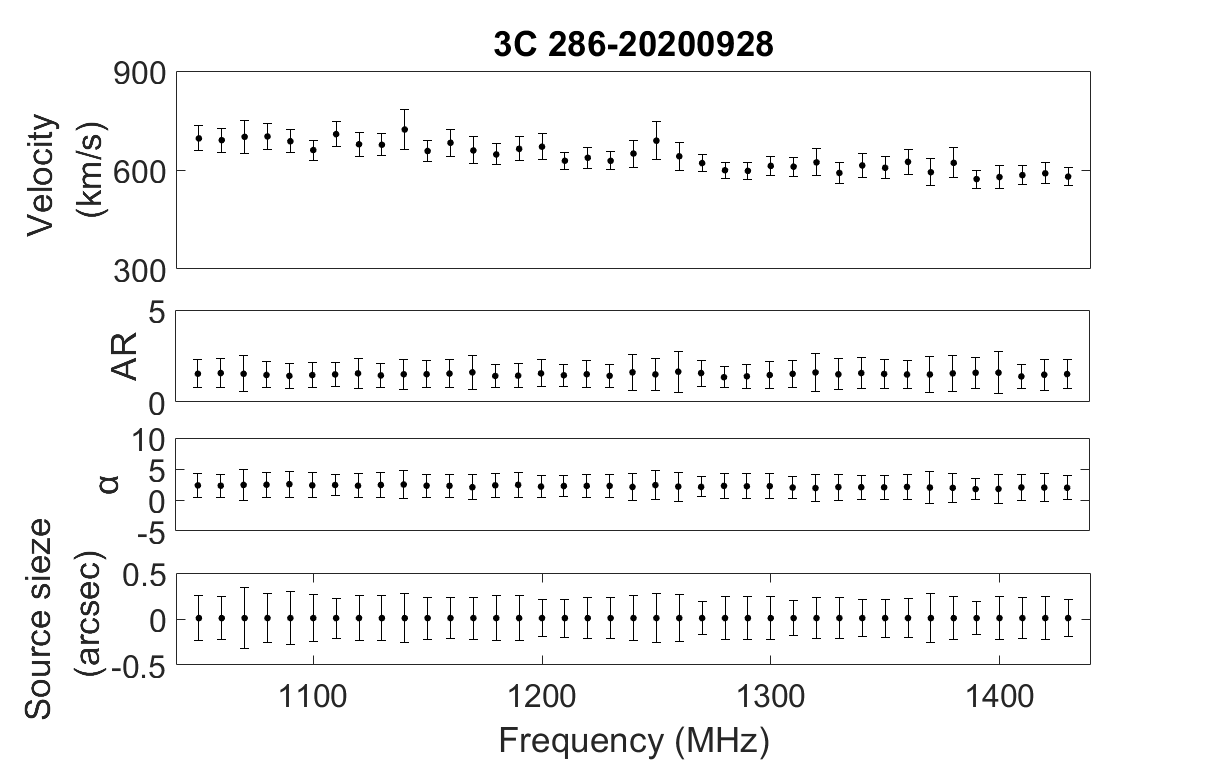}
    \caption{The solar wind parameters of source 3C\,286 resulted from a continuous 10\,MHz bandwidth sub-band data using 80\,s time length on September 28, 2020. The four fitting parameters velocity, AR, $\alpha$ and source size are displayed from top to bottom panels. The dots and solid lines represent the fitting values and the error bars (95$\%$ confidence intervals).}
    \label{parameters}

\end{figure}

In our previous studies \citep{2021MNRAS.504.5437L}, an optimized model-fitting method has been developed, which can obtain the fitting value of four solar wind parameters (V, AR, $\alpha$, $\theta$) and the fitting errors simultaneously. The parameters fitting results of our method are deterministic, which means a set of initial parameters corresponding to an identical result. We applied multi-initialization strategy \citep{article} to find the global minimum, which guarantees the fitting convergence in the global minimization. Fig.~\ref{parameters} demonstrates one of the fitting results, as an example of radio source 3C\,286\, which are resulted from a continuous 10\,MHz bandwidth sub-bands data using 80\,s time length on September 28, 2020. The dots and solid lines show the fitting values and the errors of the four parameters. \\

\section{Discussions}

In this section, a linear changing trend and a minutes scale oscillation of the solar wind velocity are presented and discussed.\\


The velocities through the whole frequency band (Channels affected by strong RFI are rejected) were fitted by a weighted linear function, which is given below in equation~(\ref{eq:2})

\begin{equation}
    V=k\times x+b
	\label{eq:2}
\end{equation}

\begin{figure}
	\includegraphics[width=\columnwidth]{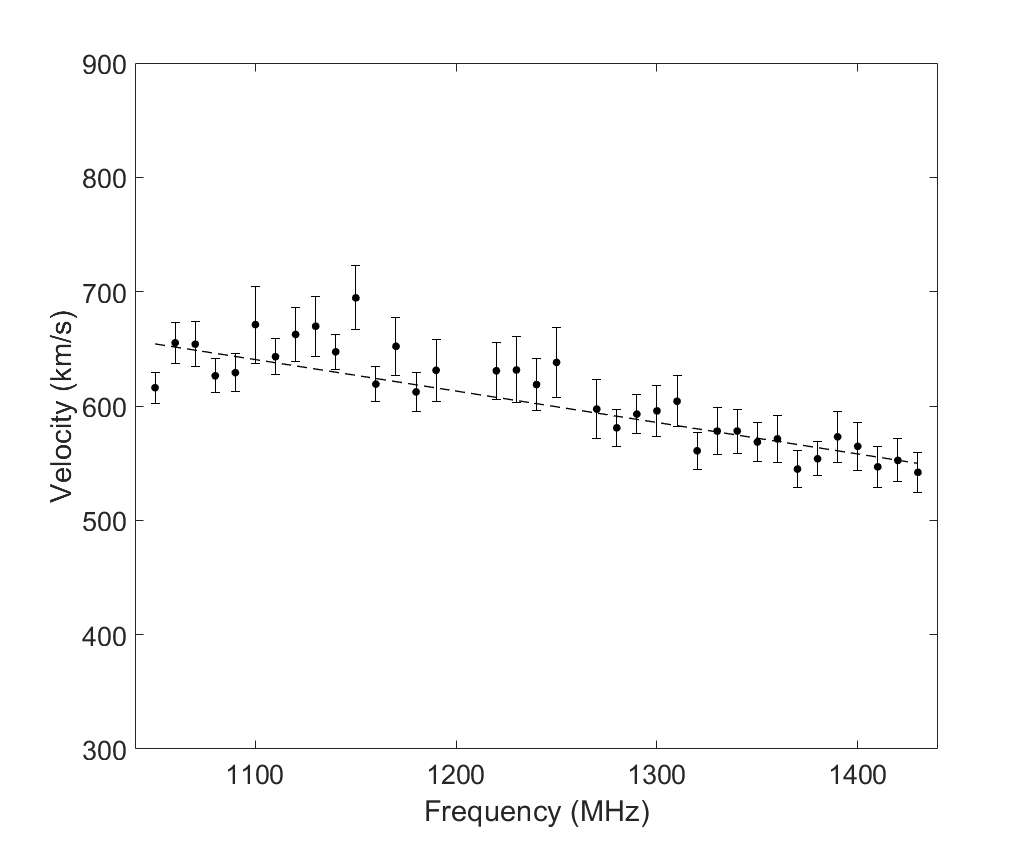}
    \caption{An example of the solar wind velocities through sub-bands. The radio source was 3C\,286 observed by FAST on September 27, 2020. The dots and the vertical lines are the fitted solar wind velocities and errors, and the dashed line shows the linear fit of the velocities with error.(Channels affected by strong RFI are rejected.)}
    \label{velocity}
\end{figure}

Where $x=1+\frac{f-1050}{10} $ is the serial number of each sub-set, $f$ is the observing frequency of each sub-band in MHz, and $V$ is the solar wind velocity in $km/s$. The dots and the vertical lines are the fitted solar wind velocities and errors, and the dashed line shows the linear fit of the velocities with error. The best fit parameters of the dashed line in Fig.~\ref{velocity} are $k=-2.8\pm0.3$ and $b=655.8\pm6.0$. The same procedure was also performed on the data observed during the three days. Totally, we got 80 sets of values for $k$ and $b$.\\

\begin{figure}
	\includegraphics[width=\columnwidth]{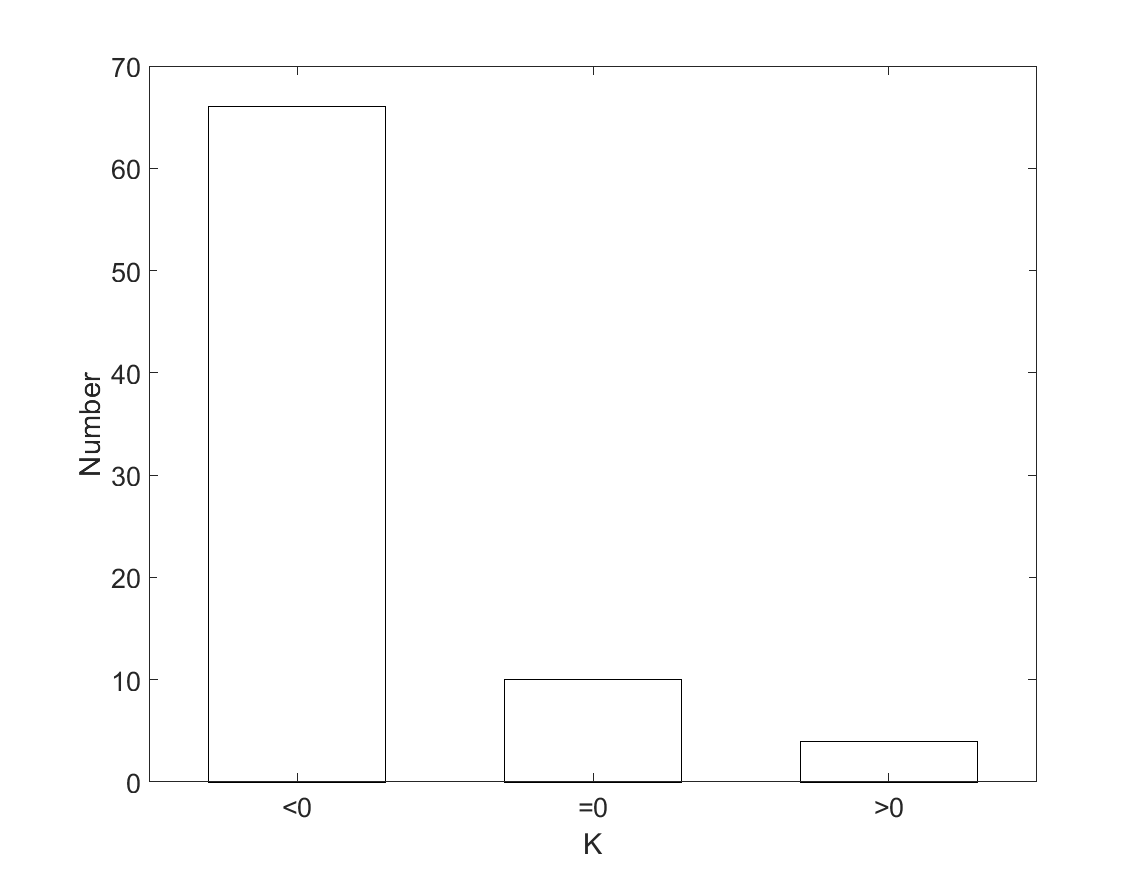}
    \caption{The distribution of $k$ during September 26-28, 2020. }
    \label{k_dis}
\end{figure}

 The framework of classical IPS theory is based on some standard assumptions, including the temporal intensity variation of a scintillating source is mainly due to the bulk motion of the solar wind across the line of sight \citep{2000Oberoi}. According to the definition of IPS, the velocity derived from IPS phenomenon reflects the bulk speed of the solar wind, therefore over the whole observing frequency range, the velocities at each frequency should be the same only with a small fluctuation. However, Fig.~\ref{velocity} clearly shows a linear trend over the whole observing frequency, which has not yet been reported in literature, and this represents one of the studies that one can make using the capabilities of FAST. The histogram in Fig.~\ref{k_dis} displayed the distribution of $k$ during September 26-28, 2020. The statistics of $k$ through the three days show that over $80\%$ of $k$ are negative. A similar phenomenon has also been seen from the observations of IPS by the Arecibo Telescope also using $L$-band (Arecibo team, private communications). \\

 There are three possible origins causing the linear trend: 1) the observing facility, 2) the model-fitting pipeline, 3) the intrinsic properties of the solar wind plasma. \\

 In order to figure out the real cause of this trend, the three possibilities were investigated one by one. After the RFI removal and the baseline flatten by the mean subtract process, the difference caused by the instrument system has been eliminated, as shown in Fig.~\ref{data_process}. So, the first possibility can be excluded. For the purpose to investigate the influence of the model-fitting pipeline, a simulation was carried out. We adopted a characteristic solar wind velocity value from our observations as the simulated one. The different S/N ratios in different frequency channels were achieved by inserting the scaled data from our off-source observation. The simulated data was processed by the same model-fitting method, which was applied to our real data. Then we can inspect the influence of the method on our result by comparing the recovered value with input value. The simulation procedure is described below:\\\\
 1. Generate a series of theoretical power spectra according to equation~(\ref{eq:1}) with a same solar wind velocity value over the observing frequency range from 1.05 to 1.45\,GHz.\\
 2. Carry out inverse FFT to the spectra to create the time domain data.\\
 3. Add the real off-source data observed by FAST to each time domain data with a linear change from 14\,dB at the lowest frequency 1.05\,GHz to 11\,dB at the highest frequency 1.45\,GHz.\\
 4. Perform the model-fitting pipeline with the time domain data to obtain the fitting values and errors.\\

\begin{figure}
	\includegraphics[width=\columnwidth]{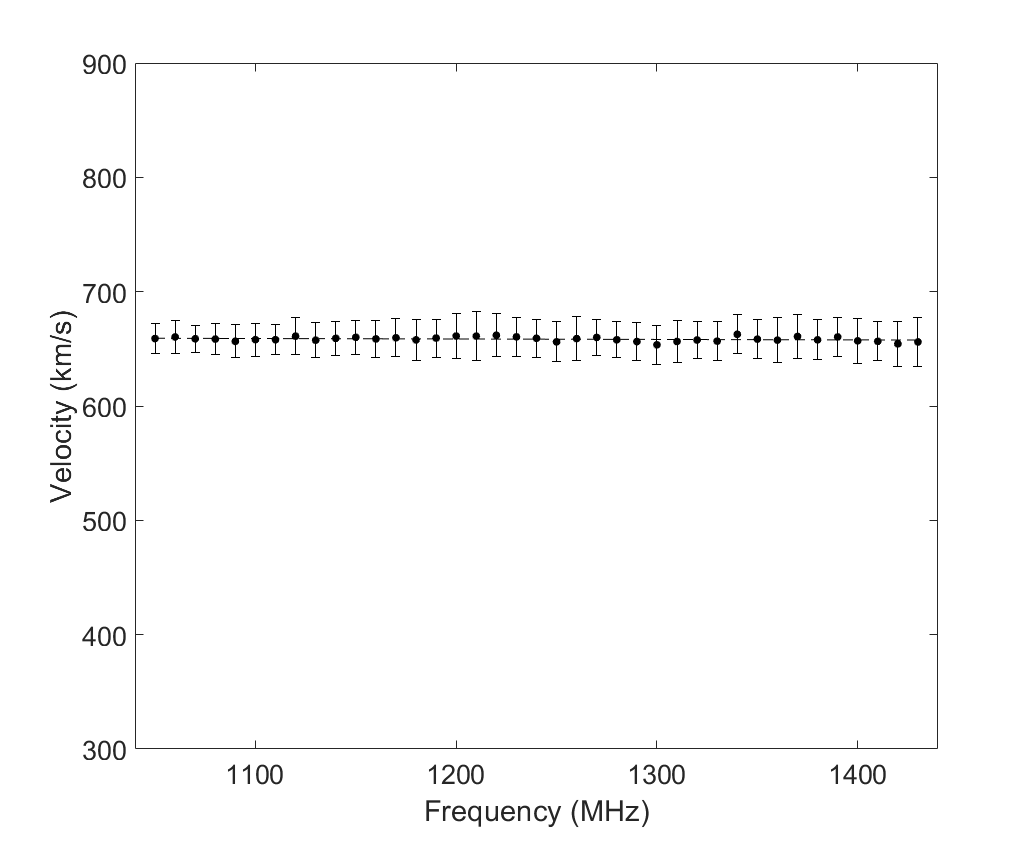}
    \caption{The linear fit result of the simulated data. The dots and solid line represent the fitting values and the errors. The dashed line in the middle shows the weighted linear fit of the velocities. }
    \label{simulation}
\end{figure}

 Fig.~\ref{simulation} displays the model-fitting results of the simulated data. The dots and solid lines represent the fitting values and the errors. The dashed line in the middle shows the linear fit with the error of the velocities. The $V$ used for simulation is $650\,km/s$ and $k=-0.04\pm0.2$. The solar wind velocities through the whole frequency range demonstrate an excellent consistency, which means the velocity trend shown in Fig.~\ref{velocity} is not caused by the model-fitting method. \\

 Both of the first two possible causes are unable to result in this velocity changing trend shown in Fig.~\ref{velocity}. So the different solar wind velocities measured in different FAST frequency channels could be caused by some intrinsic characteristics of the solar wind, requiring a theoretical work to further study the radiative transfer mechanism with a physics-based turbulent velocity field. \\

 The solar wind magnetohydrodynamic turbulence, though overall balanced at a macro scale, is always locally imbalanced in creating patches of positive and negative cross-helicities \citep{2008ApJ...672L..61P}.  The solar wind turbulence cascade could be intermittently decaying towards these Fresnel-scale irregularities, which might be ascribed to a gradual misalignment of electron density irregularities at different spatial Fresnel scales ($\sqrt{Z\lambda}$, where $Z\approx 1.2\times{10}^{11}\,m$, and $0.21\,m < \lambda  < 0.29\,m $) of about 150 to 190\,km. A big turbulence eddy associated with a large density irregularity tends to be a bit faster than its nested small eddies, resulting in higher velocities at lower frequencies, which causes linear change trend of the solar wind velocity. Such an inconsistency in cross-scale density irregularities is likely caused by a cascade of solar wind turbulence with a minutes scale.\\

\begin{figure}
\centering
    \subfigure[]{
    \includegraphics[width=\columnwidth]{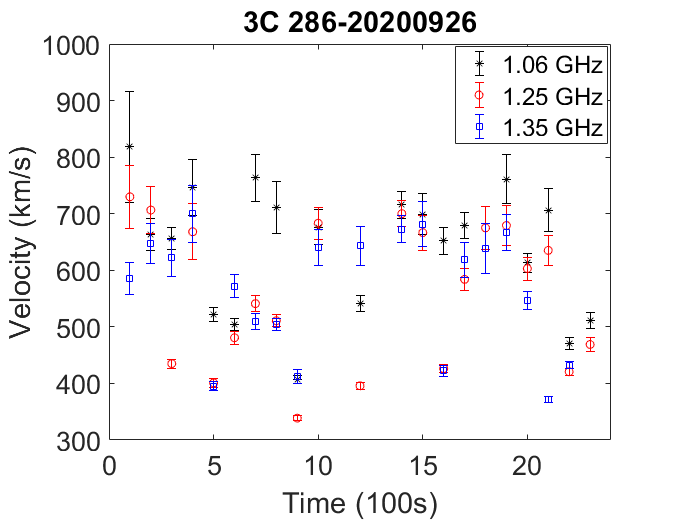}
    }
    \subfigure[]{
    \includegraphics[width=\columnwidth]{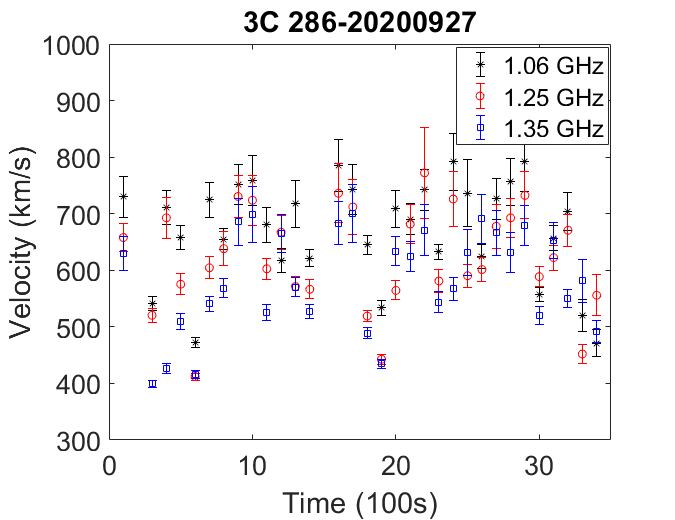}
    }
    \subfigure[]{
    \includegraphics[width=\columnwidth]{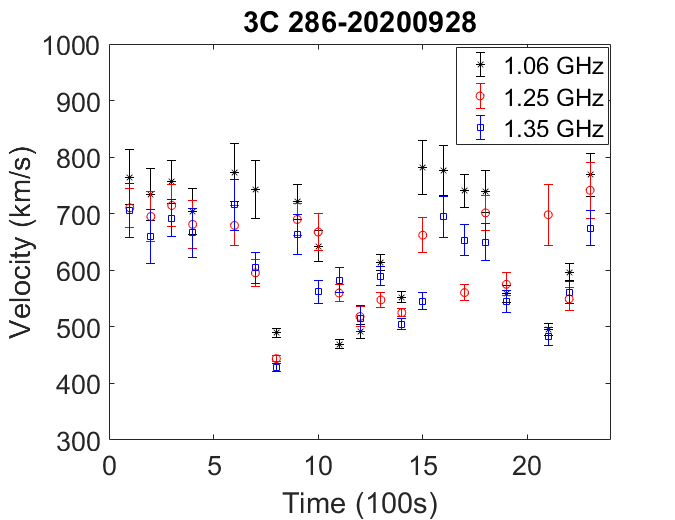}
    }
	\caption{The solar wind velocity variations through September 26-28, 2020. The black dots, red circles, blue squares and solid lines represent the velocities and the fitting errors obtained at 1.06\,GHz, 1.25\,GHz and 1.35\,GHz, respectively (the data affected by strong RFI are rejected). }
    \label{velocity_new}
\end{figure}

Since most of the solar wind velocities show a linear trend, the velocities at some certain frequencies can characterize the change in velocity over time. Fig.~\ref{velocity_new} presents the solar wind velocity change during September 26-28, 2020. The black dots, red circles, blue squares, and solid lines represent the velocities and their fitting errors obtained at 1.06\,GHz, 1.25\,GHz and 1.35\,GHz, respectively (the data affected by strong RFI are rejected). The velocities deduced from the three different frequencies have also revealed the phenomenon demonstrated in Fig.~\ref{velocity}, which is the velocity decreases with the increasing observing frequency. Furthermore, the structure function of the velocities show that there appears a 3-5 minute-scale oscillation in the solar wind velocity. The velocity oscillation may imply that, during our observation with FAST, a slow change of the background solar wind, an evidence of high- and low-speed streams, the quasi-periodic electron-density fluctuations \citep{2012AdSpR..49..500E}, or the presence of slow-mode wave in the solar wind which is caused by the parametric decay instability of Alfv\'en wave \citep{2013JGRA..118.7507T, 2018ApJ...860...17S, 2018ApJ...866...38R, 2019ApJ...880L...2S}. \\

%

\section{Concluding remarks}

In conclusion, the Five-hundred-meter Aperture Spherical radio Telescope (FAST) has outstanding advantages to carry out observations of IPS. The high sensitivity allows it to discover weak signals that can not be detected by other telescopes. The wide coverage of observing frequency makes it capable of effectively removing the RFI and performing dynamic spectrum analysis. \\

Based on the observations of IPS by FAST, we observed a linear trend of solar wind velocity change in the entire frequency band of FAST 19-beam receiver, and an oscillation in minutes scale. The linear trend represents a new result, which cannot be done by most of the existent radio telescopes. According to the FAST observation, the interplanetary solar wind flow is variable on timescale of some minutes, and the solar wind density irregularities are not strictly co-moving with the speed of its bulk flow. These observations of the radio source 3C\,286 during September 26-28, 2020 show a very exciting IPS prospect, but also revoke the necessity for more studies.\\

\section*{Acknowledgements}

This work is supported by the National Key $R \& D$ Program of China under grant number 2018YFA0404703, the Open Project Program of the CAS Key Laboratory of FAST, NAOC, Chinese Academy of Sciences, the Specialized Research Fund for State Key Laboratories, the basic research program and project of Yunnan province of China (2019FB009) as well as the National Natural Science Foundation of China under grant number 41874205. The authors thank M.M. Bisi (RAL Space, United Kingdom Research and Innovation - Science and Technology Facilities Council), for his conversations about the Arecibo comparisons. We are grateful to M. Tokumaru (Institute for Space-Earth Environmental Research, Nagoya, Japan) and Jian-Sen He (School of earth and space sciences, Peking University), for their helpful discussions. The authors also thank Dr. Juan Arratia, for his help in polishing the English. This work made use of the data from FAST (Five-hundred-meter Aperture Spherical radio Telescope). FAST is a Chinese national mega-science facility, operated by the National Astronomical Observatories, Chinese Academy of Sciences.

\section*{Data Availability}

According to the FAST data policy, the data underling this article has been released. Please contact FAST Data Center (fastdc@nao.cas.cn) for the data.



\bibliographystyle{mnras}
\bibliography{ips} 




%
%


\bsp	
\label{lastpage}
\end{document}